# Radiation pressure on a dielectric wedge

**Masud Mansuripur[†], Armis R. Zakharian[‡] and Jerome V. Moloney[†, ‡]**

[†]*Optical Sciences Center and* [‡]*Department of Mathematics, University of Arizona, Tucson, Arizona 85721*
*masud@optics.arizona.edu*

**Abstract:** The force of electromagnetic radiation on a dielectric medium may be derived by a direct application of the Lorentz law of classical electrodynamics. While the light's electric field acts upon the (induced) bound charges in the medium, its magnetic field exerts a force on the bound currents. We use the example of a wedge-shaped solid dielectric, immersed in a transparent liquid and illuminated at Brewster's angle, to demonstrate that the linear momentum of the electromagnetic field within dielectrics has neither the Minkowski nor the Abraham form; rather, the correct expression for momentum density has equal contributions from both. The time rate of change of the incident momentum thus expressed is equal to the force exerted on the wedge plus that experienced by the surrounding liquid.

## 1. Introduction

In a previous paper [1] we showed that a consistent picture of the electromagnetic momentum in dielectric media can be obtained by a direct application of the Lorentz law of force. A major conclusion was that, for a plane electromagnetic wave inside a dielectric medium, momentum density may be expressed as the average of two expressions associated with the names of H. Minkowski ($p = \frac{1}{2} D \times B$) and M. Abraham ($p = \frac{1}{2} E \times H / c^2$). Another conclusion was that the local electromagnetic force density may be obtained by adding the Lorentz force of the *E*-field on the induced bound charges to the Lorentz force of the *H*-field on the induced bound currents. We argued that, whereas bound currents are typically distributed throughout the volume of a given medium, bound charges (in isotropic, homogeneous dielectrics) appear only on the surfaces and/or interfaces with adjacent media.

An interesting manifestation of the magnetic Lorentz force appears at the edges (i.e., side-walls) of a finite-diameter beam of light, where, depending on the beam's polarization state, the host medium may experience either an expansive or a compressive force.

The present paper demonstrates the consistency of the results obtained in [1] in the special case of a wedge-shaped dielectric prism illuminated by a *p*-polarized plane-wave at Brewster's angle of incidence. The prism's apex angle is such that, at the exit facet of the wedge, the beam arrives once again at Brewster's angle. The choice of this particular geometry simplifies the analysis by suppressing all interfacial reflections. In Section 3 the total force exerted on a dielectric wedge in free-space is shown to be equal to the time rate of change of the incident optical momentum. This is reassuring, in light of the fact that the electromagnetic momentum in free-space is not the subject of any controversies [2]. The case of a dielectric wedge immersed in a transparent liquid is investigated in Section 4, where, once again, the net force is shown to be equal to the time rate of change of the incident optical momentum. This time around, however, the momentum density is assumed to be given by the average of Minkowski and Abraham expressions, both of which have been the subject of extensive debate for nearly a century [3-7].

As a practical matter, the conservation of momentum demonstrated in Section 4, fundamental as it may be, is perhaps not as interesting as the actual division of the transferred momentum between the solid object and its surrounding liquid. In Section 5 we argue for a particular method of allocating a fraction of the interfacial charge density (and the corresponding Lorentz force) to each of two adjacent dielectrics. The method is subsequently employed in Section 6 to compute the net force of radiation exerted on a solid wedge immersed in a liquid. Section 7 presents the results of our numerical simulations of a Gaussian beam incident on a dielectric prism, which confirm the analytical results of the preceding sections. Our concluding remarks appear in Section 8.

## 2. Notation and basic definitions

The MKSA system of units is used throughout this paper. To compute the force of the electromagnetic radiation on a given medium, we use Maxwell's equations to determine the distributions of the *E*- and *H*-fields (both inside and outside the medium). We then apply the Lorentz law

$$\bm{F} = \rho_b \bm{E} + \bm{J}_b \times \bm{B}, \tag{1}$$

where $\bm{F}$ is the force density, and $\rho_b$ and $\bm{J}_b$ are the bound charge and current densities, respectively [2,7]. The magnetic induction $\bm{B}$ is related to the *H*-field via $\bm{B} = \mu_o \bm{H}$, where $\mu_o = 4\pi \times 10^{-7}$ henrys/meter is the permeability of free space. In the absence of free charges $\nabla \cdot \bm{D} = 0$, where $\bm{D} = \varepsilon_o \bm{E} + \bm{P}$ is the displacement vector, $\varepsilon_o = 8.8542 \times 10^{-12}$ farads/meter is the free-space permittivity, and $\bm{P}$ is the local polarization density. In linear media, $\bm{D} = \varepsilon_o \varepsilon \bm{E}$, where $\varepsilon$ is the medium's relative permittivity; hence, $\bm{P} = \varepsilon_o(\varepsilon - 1)\bm{E}$. We ignore the frequency-dependence of $\varepsilon$ throughout the paper and treat the media as non-dispersive.

When $\nabla \cdot \bm{D} = 0$, the bound-charge density $\rho_b = -\nabla \cdot \bm{P}$ may be expressed as $\rho_b = \varepsilon_o \nabla \cdot \bm{E}$. Inside a homogeneous and isotropic medium, $\bm{E}$ being proportional to $\bm{D}$ and $\nabla \cdot \bm{D} = 0$ imply that $\rho_b = 0$; no bound charges, therefore, exist inside such media. However, at the interface between two adjacent media, the component of $\bm{D}$ perpendicular to the interface, $\bm{D}_\perp$, must be continuous, implying that $\bm{E}_\perp$ is discontinuous and, therefore, bound charges exist at such interfaces; the interfacial bound charges thus have areal density $\sigma_b = \varepsilon_o(E_{2\perp} - E_{1\perp})$. Under the influence of the local *E*-field, these charges give rise to a Lorentz force density $\bm{F} = \tfrac{1}{2} Real(\sigma_b \bm{E}^*)$, where $\bm{F}$ is the force per unit area of the interface. Since the tangential *E*-field, $\bm{E}_{||}$, is continuous across the interface, there is no ambiguity as to the value of $\bm{E}_{||}$ that should be used in computing the force. As for the perpendicular component, the average $\bm{E}_\perp$ across the boundary, $\tfrac{1}{2}(\bm{E}_{1\perp} + \bm{E}_{2\perp})$, must be used in calculating the interfacial force [1].

The only source of electrical currents within dielectric media are the oscillating dipoles, with the bound current density being $\underline{J}_b = \partial \underline{P}/\partial t = \varepsilon_o(\varepsilon - 1)\partial \underline{E}/\partial t$. Assuming time-harmonic fields with the time-dependence factor $\exp(-i\omega t)$, we can write $J_b = -i\omega\varepsilon_o(\varepsilon - 1)E$. The $B$-field of the electromagnetic wave exerts a force on the bound current according to the Lorentz law, namely, $F = \frac{1}{2} Real\,(J_b \times B^*)$, where $F$ is force per unit volume. We have shown in [1] that, among other things, this magnetic Lorentz force is responsible for a lateral pressure exerted on the host medium at the edges of a finite-diameter beam; the force per unit area at each edge (i.e., side-wall) of the beam is given by

$$F^{(edge)} = \tfrac{1}{4}\varepsilon_o(\varepsilon - 1)|E|^2. \quad (2)$$

Here $|E|$ is the magnitude of the $E$-field of a (finite-diameter) plane-wave in a medium of dielectric constant $\varepsilon$. If the $E$-field is parallel (perpendicular) to the beam's edge, the force is compressive (expansive); in other words, the opposite side-walls of the beam tend to push the medium toward (away from) the beam center. The edge force does not appear to be sensitive to the detailed structure of the beam's edge; in particular, a one-dimensional Gaussian beam exhibits the edge force described by Eq. (2) when its (magnetic) Lorentz force on the host medium is integrated laterally on either side of the beam's center [1].

Finally, the momentum density (per unit volume) of a plane wave within a dielectric medium of refractive index $n$ ($=\sqrt{\varepsilon}$) was shown in [1] to be the average of Minkowski's $\frac{1}{2} D \times B$ and Abraham's $\frac{1}{2} E \times H /c^2$, namely,

$$p = \tfrac{1}{4}(n^2 + 1)\,n\varepsilon_o|E|^2/c. \quad (3)$$

Considering that the speed of light in the (dispersionless) medium is $c/n$, the rate of flow of optical momentum per unit area per unit time is thus $\tfrac{1}{4}(n^2 + 1)\varepsilon_o|E|^2$. These results will be used in the following sections to establish the equivalence of force and the time rate of change of incident optical momentum on a dielectric prism.

## 3. Dielectric prism illuminated at Brewster's angle

In this section we compute the force exerted by a (finite-diameter) plane wave on a dielectric prism residing in free space, and show that this force is exactly equal to the time rate of change of the incident optical momentum. With reference to Fig. 1, a $p$-polarized plane wave is incident on a dielectric prism of apex angle $\phi$ and refractive index $n$ at Brewster's angle $\theta_B$ ($\tan\theta_B = n$). The refracted angle inside the slab is given by $\tan\theta'_B = 1/n$. The apex angle is $\phi = 2\theta'_B$, so the internal angle of incidence on the exit facet is also equal to Brewster's angle. Since the reflectivity at Brewster's angle is zero, the only beams in this system are the incident beam, the refracted beam inside the prism, and the transmitted beam. Inside the prism, the $H$-field is the same as that outside, as required by the continuity of $H_\parallel$ at the interfaces. Similarly, the continuity of $E_\parallel$ and $D_\perp$ at the interfaces require that, inside the prism and just beneath the surface, $E_\parallel = E_o\cos\theta_B$, and $E_\perp = (E_o/n^2)\sin\theta_B$, which fixes the magnitude of $E$ inside the prism at $E_o/n$.

Inside the prism, the bound-charge density $\rho_b$ is zero, and the magnetic field of the light everywhere (except at the beam's edges) is 90° out of phase relative to the bound-current density $J_b$. The electric component of the Lorentz force is due to the bound charges induced on the entrance and exit facets, the density of which is obtained from the $\varepsilon_o E_\perp$ discontinuity, namely,

$$\sigma_b = \varepsilon_o E_o(1 - 1/n^2)\sin\theta_B. \quad (4)$$

In the direction parallel to the surface, $E_\parallel = E_o\cos\theta_B$ is continuous across the boundary, and the Lorentz force on the induced surface charges is

$$F_\| = \tfrac{1}{2}\sigma_b E_\| = \tfrac{1}{2}\varepsilon_o E_o^2 \,(1 - 1/n^2)\, \sin\theta_B \cos\theta_B. \tag{5}$$

The factor ½ accounts for time-averaging over one period of oscillations. Denoting by $a$ the footprint area of the beam on each facet of the prism, the projection of $F_\|$ on the $x$-axis (accounting for both facets) is given by

$$F_{\|x} = a\varepsilon_o E_o^2 \,(1 - 1/n^2)\, \sin^2\theta_B \cos\theta_B. \tag{6}$$

Next, we consider the perpendicular force $F_\perp$ on the top surface of the prism. Averaging $E_\perp$ just above and just below the surface, we find

$$F_\perp = \tfrac{1}{2}\sigma_b E_\perp = \tfrac{1}{4}\varepsilon_o E_o^2 (1 - 1/n^2)(1 + 1/n^2)\, \sin^2\theta_B. \tag{7}$$

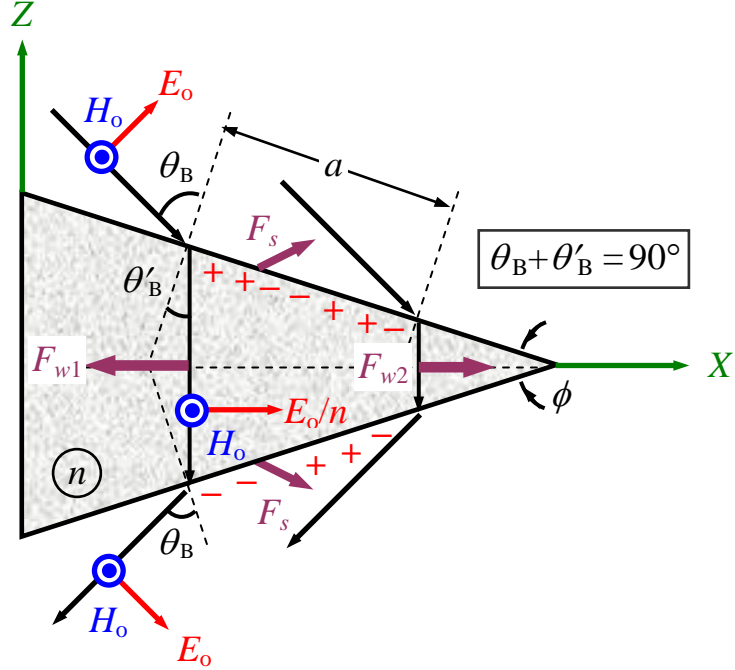

Fig. 1. Dielectric prism of index $n$, illuminated with a $p$-polarized plane wave at Brewster's angle $\theta_B$. The prism's apex angle is twice the refracted angle $\theta'_B$, and the beam's footprint on the prism's face is $a$. The $H$-field magnitude inside the slab is the same as that outside, but the $E$-field inside is reduced by a factor of $n$ relative to the outside field. The bound charges on the upper and lower surfaces feel the force $F_s$ of the $E$-field. The net force experienced by the prism is along the $x$-axis, being the sum of the forces on the upper and lower surfaces as well as $F_{w1}$ and $F_{w2}$, which are exerted at the beam's edges (i.e., side-walls) within the dielectric.

Integrating over the footprint of the beam (area = $a$), projecting onto the $x$-axis, and adding the forces on the top and bottom facets, the contribution of $E_\perp$ to the total force is found to be

$$F_{\perp x} = \tfrac{1}{2}\, a\varepsilon_o E_o^2 (1 - 1/n^4)\, \sin^2\theta_B \cos\theta_B. \tag{8}$$

Inside the prism, the force density at the side-walls of the beam, $F_w = \tfrac{1}{4}\varepsilon_o(n^2 - 1)|E|^2$, is normal to the side-walls and expansive in this case of $p$-polarized light. Here the left wall area is larger than the right wall area by $2a\sin\theta'_B$, so the net force on the side-walls, directed along the negative $x$-axis, is

$$F_{w1} + F_{w2} = -\tfrac{1}{2}\, a\, \varepsilon_o E_o^2\, (1 - 1/n^2)\, \sin\theta'_B. \tag{9}$$

Adding the force components in Eqs. (6), (8), and (9), we find the net force on the prism to be

$$F_x^{(total)} = (a\cos\theta_B)\varepsilon_o E_o^2 [(n^2 - 1)/(n^2 + 1)]. \quad (10)$$

Considering that in a beam of amplitude $E_o$ and cross-sectional area $a\cos\theta_B$ momentum arrives at the rate of $\tfrac{1}{2}(a\cos\theta_B)\varepsilon_o E_o^2$ per second, and that in the system of Fig. 1 the incoming and outgoing momenta make an angle of $2\theta'_B$ with the x-axis, where $\cos(2\theta'_B) = (n^2 - 1)/(n^2 + 1)$, the total force $F_x$ in Eq. (10) is seen to be equal to the time rate of change of the incident optical momentum.

## 4. Wedge immersed in a liquid

In a majority of optical trapping experiments the objects of interest are immersed in water or some other transparent liquid [8,10]. In this section we consider the case of a dielectric prism immersed in a liquid of lower refractive index, and determine the contribution of the Lorentz force experienced by the liquid to the total force exerted by the incident beam. Once again, the total force turns out to be equal to the time rate of change of the incident optical momentum, where the momentum density (in the liquid) is given by Eq. (3).

Figure 2 shows a dielectric prism of refractive index $n_2$ immersed in a liquid of index $n_1$. The Brewster angle in this case is given by $\tan\theta_B = n_2/n_1$ and, as before, $\theta_B + \theta'_B = 90°$ and the apex angle of the prism is $\phi = 2\theta'_B$. Keeping the same incident E-field amplitude $E_o$ as before, the previous H-field magnitude $H_o$ will have to be multiplied by $n_1$, and the E-field magnitude inside the prism becomes $n_1 E_o/n_2$. We continue to assume the same footprint, $a$, for the beam at the entrance and exit facets of the prism. The density of bound charges at the interface between the prism and the liquid may be written as follows:

$$\sigma_b = \varepsilon_o [1 - (n_1^2/n_2^2)] E_o \sin\theta_B. \quad (11)$$

Unlike the case of a dielectric prism in free space, where $\sigma_b$ was solely due to the bound charges on the exterior facets of the prism, in the present case the interfacial charge is the superposition of two adjacent and oppositely charged layers. One such layer belongs to the solid dielectric, the other to the liquid. A simple way to visualize these layers is to imagine the existence of a narrow gap around the wedge, separating the solid from its liquid host. The continuity of $D_\perp$ at the interface requires the existence of an $E_\perp$ within the gap itself. Thus the $E_\perp$ discontinuity on one side of the gap yields the charge density on the solid surface, while the corresponding discontinuity on the other side of the gap determines the charge density on the liquid surface. Presently we are concerned only with the net force on the interfacial charges and, therefore, ignore the detailed composition of the charged layer at the solid-liquid interface. In the following sections we shall return to computing the force on the solid object alone, at which point due attention will be paid to the composition of the charged layer.

The force components parallel and perpendicular to each solid-liquid interface are obtained by a method similar to that discussed in Section 3. The combined force on the upper and lower facets of the prism is along the x-axis, and its magnitude is given by

$$F_x^{(surface)} = a\varepsilon_o(1 - n_1^2/n_2^2)[1 + \tfrac{1}{2}(1 + n_1^2/n_2^2)] E_o^2 \sin^2\theta_B \cos\theta_B. \quad (12)$$

Inside the prism, the net force exerted at the beam's edges and directed along the negative x-axis is given by

$$F_{w1} + F_{w2} = -\tfrac{1}{2} a\varepsilon_o [n_1^2 - (n_1^2/n_2^2)]E_o^2 \sin\theta'_B. \quad (13)$$

In addition to the above forces, the edges of the incident and transmitted beams exert a force $F_w$ on the host liquid. As shown in Fig. 2, the uncompensated area of each edge within the liquid is $a\sin\theta_B$, and the angle between $F_w$ and the z-axis is $2\theta'_B$; therefore,

$$F_x^{(liquid)} = 2F_w \sin 2\theta'_B = \tfrac{1}{2} a \sin\theta_B (n_1^2 - 1)\varepsilon_o E_o^2 \sin 2\theta'_B. \tag{14}$$

Adding up all the forces in Eqs. (12), (13), and (14) yields

$$F_x^{(total)} = \tfrac{1}{2} a \cos\theta_B (n_1^2 + 1)\varepsilon_o E_o^2 \cos 2\theta'_B. \tag{15}$$

Considering that $a\cos\theta_B$ is the cross-sectional area of the incoming and outgoing beams, that the rate of flow of optical momentum per unit area per unit time is $\tfrac{1}{4}(n_1^2 + 1)\varepsilon_o E_o^2$, and that the incoming and outgoing momenta make an angle $2\theta'_B$ with the $x$-axis, it is clear that the total $F_x$ exerted on the media (prism plus liquid) is equal to the time rate of change of the incident optical momentum.

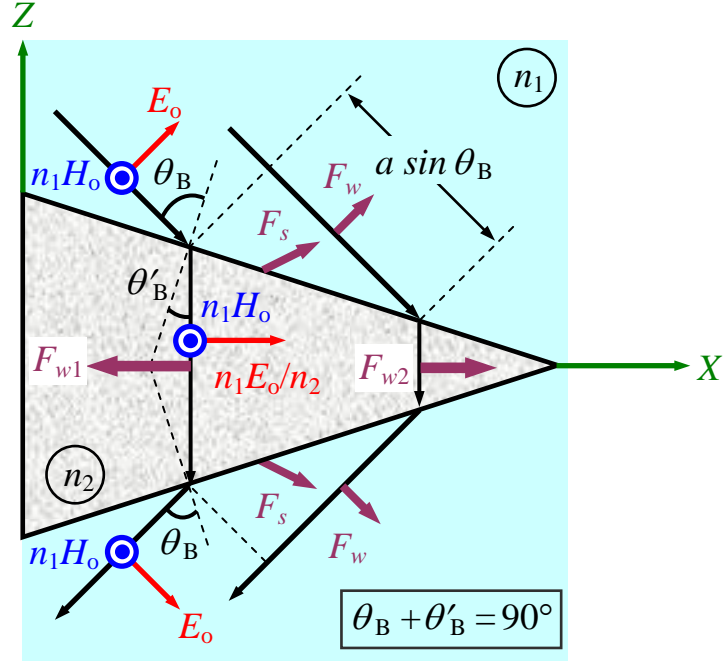

Fig. 2. Dielectric prism of index $n_2$ in a host liquid of index $n_1$, illuminated with a $p$-polarized plane wave at Brewster's angle $\theta_B$. The apex angle is $2\theta'_B$, and the footprint of the beam on each side of the wedge is $a$. The liquid experiences a force $F_w$ at the edges of the beam where the (expansive) force on one edge is not compensated by the force on the opposite edge. The net force exerted on the system is along the $x$-axis and is the sum of the forces $F_s$ on the upper and lower solid-liquid interfaces, $F_{w1}$ and $F_{w2}$ at the beam's edges within the prism, and $F_w$ at the beam's edges inside the liquid.

Ashkin [11] has used the equality between the imparted force and the change in optical momentum to derive the conditions under which spherical particles immersed in a liquid can be trapped by a focused laser beam. In contrast to the results of the present section, Ashkin assumes that the light rays carry the Minkowski momentum, and that any change of the momentum of the rays in their interaction with a particle is fully taken up by the particle itself. We believe the ray's momentum should be halfway between Minkowski and Abraham values, and that the contribution of incident momentum to the liquid should not be ignored.

### 5. Nature of the force exerted on interfacial charged layer

To describe the different roles played by the interfacial charges belonging to the solid and liquid sides of the interface, we assume the presence of a small gap – much shorter than the wavelength $\lambda_o$ of the light – at the interface between the liquid of dielectric constant $\varepsilon_l$ and solid of dielectric constant $\varepsilon_2$ shown in Fig. 3. If the perpendicular $E$-field in the gap is denoted by $E_g$, the continuity of $\mathbf{D}_\perp$ yields the amplitude of $\mathbf{E}_\perp$ in the two media as $E_g/\varepsilon_1$ and $E_g/\varepsilon_2$, as indicated. The surface charge densities are then given by the $\mathbf{E}_\perp$ discontinuity at each interface, namely, $\sigma_1 = \varepsilon_0 [(1/\varepsilon_1) - 1]E_g$ and $\sigma_2 = \varepsilon_0 [1 - (1/\varepsilon_2)]E_g$. In the limit when the gap shrinks to zero, the net charge density becomes $\sigma = \sigma_1 + \sigma_2 = \varepsilon_0 [(E_g/\varepsilon_1) - (E_g/\varepsilon_2)]$, as expected. The effective $\mathbf{E}_\perp$ acting on each charge layer (thus exerting a force on the corresponding medium) is the average $E$-field at the boundary. Therefore, the effective (perpendicular) force per unit area acting on $\sigma_1$ is $F_1 = ¼\sigma_1 [(1/\varepsilon_1) + 1]E_g$, which, upon substitution for $\sigma_1$, becomes

$$F_1 = ¼\varepsilon_0 [(1/\varepsilon_1)^2 - 1]\,|E_g|^2. \tag{16}$$

Similarly, the effective, time-averaged perpendicular force on $\sigma_2$ is

$$F_2 = ¼\varepsilon_0 [1 - (1/\varepsilon_2)^2]\,|E_g|^2. \tag{17}$$

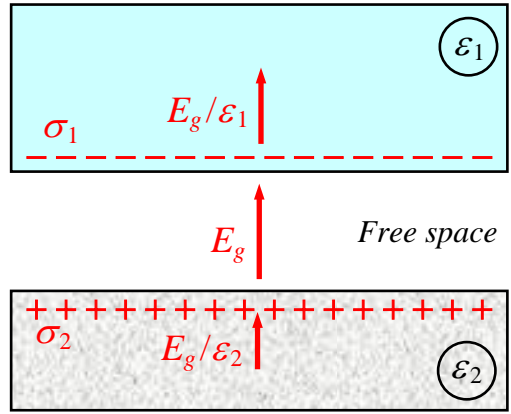

Fig. 3. A small gap at the interface between a solid (dielectric constant = $\varepsilon_2$) and a liquid (dielectric constant = $\varepsilon_1$) helps to explain the existence of separate interfacial charges. If the perpendicular $E$-field in the gap is denoted by $E_g$, the continuity of $\mathbf{D}_\perp$ across the gap yields the magnitude of $\mathbf{E}_\perp$ in the two media as $E_g/\varepsilon_1$ and $E_g/\varepsilon_2$, as shown. The surface charge densities are then given by the $\mathbf{E}_\perp$ discontinuity at each interface.

Upon shrinking the gap to zero, the net force acting on the interfacial charge layer $\sigma$ turns out to be $F = F_1 + F_2 = ¼\varepsilon_0 [(1/\varepsilon_1)^2 - (1/\varepsilon_2)^2]\,|E_g|^2$, which is identical to the total charge density $\sigma$ times the net (effective) $E$-field $E_{\text{eff}} = ½[(E_g/\varepsilon_1) + (E_g/\varepsilon_2)]$ at the solid-liquid interface. Note, however, that $F_1$ and $F_2$ *cannot* be obtained by simply multiplying this $E_{\text{eff}}$ into $\sigma_1$ and $\sigma_2$, respectively. The reason is that an additional (attractive) Coulomb force exists between the two charged layers, namely,

$$F_\text{o} = ±¼\sigma_1\sigma_2/\varepsilon_0 = ±¼\varepsilon_0 [1 - (1/\varepsilon_1)][1 - (1/\varepsilon_2)]\,|E_g|^2. \tag{18}$$

The correct expressions for $F_1$ and $F_2$ emerge when the above $F_\text{o}$ is added to $\sigma_1 E_{\text{eff}}$ and $\sigma_2 E_{\text{eff}}$, respectively.

It is interesting to note the special case of $\varepsilon_1 = \varepsilon_2 = \varepsilon \neq 1$. In the absence of the gap, $E_\perp$ is continuous and there are no net charges, nor forces, at the boundary. When the gap is introduced, however, $E_g = \varepsilon E_\perp$, $\sigma_2 = -\sigma_1 = \varepsilon_0(\varepsilon - 1)E_\perp$ and $F_2 = -F_1 = ¼\varepsilon_0(\varepsilon^2 - 1)|E_\perp|^2$. The net charge and the net force are still zero, of course, but each medium experiences a force arising from two sources: the force of $E_\perp$ on the corresponding interfacial charge density, given by $±½\varepsilon_0(\varepsilon - 1)|E_\perp|^2$, and the force $F_\text{o}$ of Eq. (18) between the two charged layers, given by $±¼\varepsilon_0(\varepsilon - 1)^2|E_\perp|^2$. For each medium, the two contributions to the force are in the same direction and thus reinforce each other. Note that the parallel component $E_\parallel$ of the $E$-field, if any, will exert on each medium a corresponding force $\pm\tfrac{1}{2}\varepsilon_o(\varepsilon-1)\mathit{Real}\,(E_\perp E_\parallel^*)$ in the direction parallel to the junction. These are quite strong forces whose existence is masked by the continuity of $\varepsilon$ at the boundary, but brought out clearly when a gap is imagined to exist between the two media.

## 6. Force experienced by prism immersed in a liquid

The net force experienced by the prism of Section 4 may now be computed by adding the internal forces exerted at the beam's edges to those exerted on the accumulated charges at the entrance and exit facets of the prism. The (magnetic) Lorentz force on the interior parts of the prism is given by Eq. (13), but the electric force on each surface has to be computed for the fraction of the charge that appears on the solid side of the interface, as explained in Section 5.

The parallel $E$-field at the solid-liquid interface is $E_\parallel = E_o\cos\theta_B$, which, by virtue of its continuity, is the same in the solid, liquid, and the gap. The perpendicular $E$-field on the liquid side of the interface is $E_\perp = E_o\sin\theta_B$, which leads to a gap field of $E_g = \varepsilon_1 E_o\sin\theta_B$. The surface charge density on the solid side of the interface is thus given by

$$\sigma_2 = \varepsilon_o[1-(1/\varepsilon_2)]E_g = \varepsilon_o[\varepsilon_1-(\varepsilon_1/\varepsilon_2)]E_o\sin\theta_B. \tag{19}$$

The force of $E_\parallel$ on these charges, when projected onto the $x$-axis and multiplied by $2a$ (to account for the illuminated area on both facets of the prism) is

$$F_{\parallel x} = a\sigma_2 E_\parallel \cos\theta'_B = a\varepsilon_o[\varepsilon_1-(\varepsilon_1/\varepsilon_2)]E_o^2\sin^2\theta_B\cos\theta_B. \tag{20}$$

As for the contribution of $E_\perp$ to the force exerted on the prism, we substitute for $E_g$ in Eq. (17), then multiply by $2a\cos\theta_B$ to account for the illuminated areas on both sides of the wedge as well as for projection onto the $x$-axis. We thus find

$$F_{\perp x} = \tfrac{1}{2}a\varepsilon_o[\varepsilon_1^2-(\varepsilon_1/\varepsilon_2)^2]E_o^2\sin^2\theta_B\cos\theta_B. \tag{21}$$

The total force on the prism is the sum of the forces in Eqs. (13), (20) and (21). Given that $\sin^2\theta_B = \varepsilon_2/(\varepsilon_1+\varepsilon_2)$, we obtain

$$F_x^{(solid)} = \tfrac{1}{2}(a\cos\theta_B)\varepsilon_o E_o^2\,\varepsilon_1(\varepsilon_1+1)(\varepsilon_2-1)/(\varepsilon_1+\varepsilon_2). \tag{22}$$

Note that for both the incident and transmitted beams, the momentum flow rate (per unit area per unit time) is $\tfrac{1}{4}\varepsilon_o(\varepsilon_1+1)E_o^2$, the cross-sectional area is $a\cos\theta_B$, and the angle between the momentum vector and the $x$-axis is $2\theta'_B$, where $\cos 2\theta'_B = (\varepsilon_2-\varepsilon_1)/(\varepsilon_2+\varepsilon_1)$. The time rate of change of the incident momentum must, therefore, be multiplied by $\varepsilon_1(\varepsilon_2-1)/(\varepsilon_2-\varepsilon_1)$ to yield the net force on the prism. That this coefficient exceeds unity (when $\varepsilon_1 > 1$) should not be surprising considering that the force of radiation on the body of the liquid is directed along the negative $x$-axis, and is given by

$$F_x^{(liquid)} = -\tfrac{1}{2}(a\cos\theta_B)\varepsilon_o E_o^2\,\varepsilon_2(\varepsilon_1^2-1)/(\varepsilon_1+\varepsilon_2). \tag{23}$$

The sum of the forces in Eqs.(22) and (23) is equal to the time rate of change of the incident momentum, as it should be. Interesting questions remain, however, as to the relevant force experienced by the prism, and as to how one might distinguish the force on the liquid from that on the solid object. By imagining a small gap at the solid-liquid interface, we have managed to disentangle these forces in a way that is rigorous and consistent with Maxwell's equations. The radiation pressure on the liquid, of course, sets in motion a hydrodynamic chain of events that redistributes the pressure within the volume of the liquid [3], and gives rise to (static and dynamic) viscoelastic forces whose investigation is beyond the scope of the present paper.

For a qualitative analysis of the "effective" force acting on the prism, it seems reasonable to assume that the magnetic Lorentz force, given by Eq. (13), and the component $F_{\|x}$ of the electric Lorentz force, given by Eq. (20), are active participants in pushing the solid object around. The role of $F_{\perp x}$ as given by Eq. (21), however, may be questioned on the grounds that the mutual attraction between the adjacent charged layers (denoted by $\sigma_1$ and $\sigma_2$ in Fig. 3) cannot possibly contribute to the prism's motion through its liquid environment. This is especially true if the two charged layers are found to intermix at the (atomically) rough exterior surface of the solid. It thus seems appropriate to remove from $F_{\perp x}$ the contribution of the attractive force between the two charged layers on each facet of the prism. This is done by using the average $E_\perp$ at the solid-liquid boundary, namely, $<E_\perp> = \frac{1}{2}[1 + (\varepsilon_1/\varepsilon_2)] E_o \sin\theta_B$, when computing the perpendicular force on the charge density $\sigma_2$ that appears on the solid side of the interface. We then find

$$\hat{F}_{\perp x} = \frac{1}{2} a\varepsilon_o [\varepsilon_1 - (\varepsilon_1/\varepsilon_2)] [1 + (\varepsilon_1/\varepsilon_2)] E_o^2 \sin^2\theta_B \cos\theta_B. \tag{24}$$

The total force on the prism is now the sum of the forces in Eqs. (13), (20) and (24), namely,

$$\hat{F}_x^{(solid)} = (a \cos\theta_B)\varepsilon_o E_o^2 \varepsilon_1(\varepsilon_2 - 1)/(\varepsilon_1 + \varepsilon_2). \tag{25}$$

Once again, the force $\hat{F}_x$ on the prism is greater than the time rate of change of the incident momentum, this time by a factor of $2\varepsilon_1(\varepsilon_2 - 1)/[(\varepsilon_1 + 1)(\varepsilon_2 - \varepsilon_1)]$. That this coefficient exceeds unity (when $\varepsilon_1 > 1$) may be understood in light of the fact that the force of radiation on the body of the liquid – again with the attractive force $F_o$ between the adjacent charged layers removed – is directed along the negative $x$-axis, and is given by

$$\hat{F}_x^{(liquid)} = -\frac{1}{2}(a \cos\theta_B)\varepsilon_o E_o^2 (\varepsilon_1 - 1). \tag{26}$$

As expected, the sum of the forces in Eqs. (25) and (26) is equal to the time rate of change of the incident momentum. We mention in passing that the force of radiation at the beam's edge on the liquid, $F_w$, when combined with the force of $E_\|$ on the surface charge density $\sigma_1$ at the liquid side of the interface, has no component on the $x$-axis. The net force of the radiation on the body of the liquid, whether the attractive force $F_o$ between the adjacent charged layers is included, as in Eq. (23), or excluded, as in Eq. (26), is therefore due solely to the action of $E_\perp$ on the surface charge density $\sigma_1$.

If the (rather extreme) position is taken that the "effective" perpendicular force on the prism surface is the force of $E_\perp$ on the *net* charge accumulated at the interface (i.e., $\sigma_1 + \sigma_2$), then the *net* force on the body of the liquid will be zero, and the force on the prism becomes

$$\hat{F}_x^{(solid)} = \frac{1}{2}(a \cos\theta_B)\varepsilon_o E_o^2 (\varepsilon_1 + 1)(\varepsilon_2 - \varepsilon_1)/(\varepsilon_2 + \varepsilon_1). \tag{27}$$

This force is exactly equal to the time rate of change of the incident optical momentum. Whether the actual force experienced by the prism is in fact given by Eq. (22), Eq. (25), or Eq. (27) is a question that cannot be decided by theoretical considerations alone; one must resort to accurate measurements of the force of radiation on the prism as a function of $\varepsilon_1$ and $\varepsilon_2$ in order to settle the question.

### 7. Computer simulations

The results of a Finite Difference Time Domain (FDTD) computer simulation of a one-dimensional Gaussian beam ($\lambda_o = 0.65$ μm) illuminating a prism of (relative) index $n = 1.5$ are shown in Fig. 4. The incidence is at Brewster's angle $\theta_B = 56.31°$, and the apex angle of the prism is $\phi = 2\theta'_B = 67.38°$.

The Gaussian beam's *H*-field profile at the outset is $H_y(x, z = 0) = H_o \exp[-(x/x_o)^2]$, where $H_o = 2.6544$ A/m and $x_o = 3.9$ μm. A time-snapshot of this *H*-field is shown in Fig. 4(a), where the beam is seen to propagate along the negative *z*-axis, enter the prism, then re-emerge into the free-space without any significant reflection losses at the boundaries. In free space, the initial *E*-field amplitude at the center of the Gaussian beam is $E_o = Z_o H_o = 1000$ V/m, where $Z_o = \sqrt{\mu_o/\varepsilon_o}$ is the impedance of free space; the magnitude of the *E*-field as it propagates along the *z*-axis and passes through the prism is depicted in Fig. 4(b). Inside the prism, the beam is seen to broaden by a factor of *n*, while its *E*-field amplitude decreases by the same factor. Ignoring for the moment the complication that the *E*-field consists of both $E_x$ and $E_z$, we approximate the *z*-component of the initial beam's Poynting vector as follows:

$$S_z(x, z = 0) = \tfrac{1}{2} E_o H_o \exp[-2(x/x_o)^2]. \tag{28}$$

The incident beam's integrated power (per unit length along the *y*-axis) is then found to be

$$\int S_z(x, z = 0)\mathrm{d}x = \sqrt{\pi/8}\, E_\mathrm{o} H_\mathrm{o}\, x_\mathrm{o} = 6.49 \times 10^{-3} \text{ W/m}. \tag{29}$$

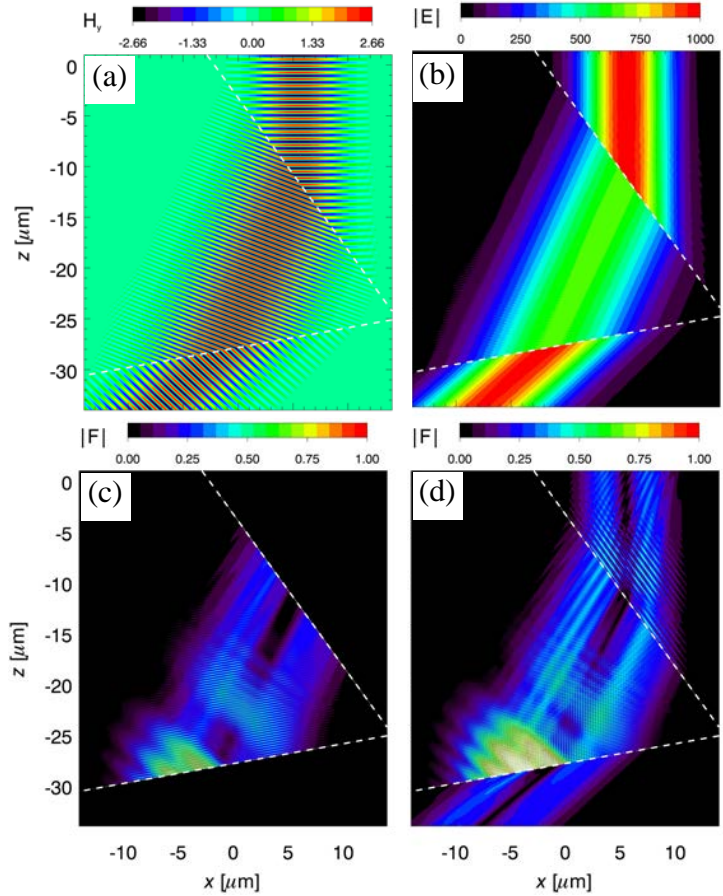

Fig. 4. Gaussian beam ($\lambda_o = 0.65$ μm, FWHM = 6.5 μm at $z = 0$) is incident at $\theta_B = 56.31°$ onto a prism of (relative) refractive index $n = 1.5$ and apex angle $\phi = 67.38°$. Since the beam is *p*-polarized and incidence is at Brewster's angle, no significant reflections occur at either the first or the second surface. In (a)-(c) the prism is in free space, whereas in (d) it is in water. The computational cell size in all cases is $\Delta x = \Delta z = 5$ nm. (a) Time-snapshot of the *H*-field, which consists only of the $H_y$ component. (b) Magnitude of the *E*-field, which consists of $E_x$ and $E_z$ components. (c) Force density distribution inside the prism of index $n = 1.5$ surrounded by free space. (d) Force density distribution inside and outside a prism of index $n_2 = 1.995$ immersed in water ($n_1 = 1.33$).

When the incident beam's environment is a liquid of refractive index $n_1$, we multiply the $H$-field by $n_1$, thus maintaining the $E$-field strength at $E_\mathrm{o} = 1000$ V/m. The integrated power of the incident beam in its liquid environment will then be $6.49\, n_1 \times 10^{-3}$ W/m.

The distribution of force density $|\boldsymbol{F}|$ in the prism of index $n = 1.5$ surrounded by free space is shown in Fig. 4(c). The plot clearly shows the magnetic Lorentz force at the edges of the beam inside the prism. The accumulated surface charges give rise to the electric Lorentz force, but their (highly localized) strength is too great to be shown on the same color scale, which has therefore been truncated in Fig. 4(c). The force density plot also shows interference fringes between the beam that arrives at the exit facet of the prism and the (residual) reflected beam. The reflection results from the rather small diameter of the beam, which corresponds to an angular spectrum that contains incidence angles in the vicinity of Brewster's angle. (The limited FDTD mesh size does not allow the simulation of a beam with a larger cross-section, hence the presence of a range of incidence angles that differ slightly from Brewster's angle is unavoidable.) The residual reflection is too weak to visibly affect the $H$- and $E$-field plots of Figs. 4(a, b), but manifests itself in the force plot of Fig. 4(c) near the exit facet of the prism. The interference fringes thus produced exhibit localized forces that oscillate back and forth between opposite directions and, therefore, cancel each other out when integrated over the entire volume of the prism. The computed total force in Fig. 4(c) is $F_x = 16.64$ pN/m, in excellent agreement with the theoretical value of $F_x = 16.58$ pN/m obtained from Eq. (10) with $(a \cos\theta_\mathrm{B})$ replaced with $\int |E(x, z = 0)/E_\mathrm{o}|^2 \mathrm{d}x = \sqrt{\pi/2}\, x_\mathrm{o}$.

Figure 4(d) shows the force density distribution in the case of a prism immersed in water ($n_1 = 1.33$, $n_2 = 1.995$, $H_\mathrm{o} = 3.53$ A/m, incident power $= 8.63 \times 10^{-3}$ W/m). Here $\theta_\mathrm{B}$ and $\theta'_\mathrm{B}$ remain the same as in the previous case, since the relative index of the prism, $n_2/n_1$, has not changed. Visible in Fig. 4(d) are interference fringes formed between the incident beam and the residual reflections at both facets of the prism, although the contribution of these fringes to the overall force remains negligible. From Eq. (15) the total force must be ½$(n_1^2 + 1)$ times the total force in the case depicted in Fig. 4(c), that is, $F_x = 22.95$ pN/m. The simulated force on the entire system (prism plus water) is found to be $F_x = 22.82$ pN/m, in excellent agreement with the aforementioned theoretical value.

## 8. Concluding remarks

In this paper we analyzed the radiation pressure on a dielectric prism in the special case where the beam arrives on both facets of the prism at Brewster's angle, thereby eliminating the complicating effects of interfacial reflections. When the surrounding medium is free space, we showed that the Lorentz force of the electromagnetic field on the induced (bound) charges and currents is exactly equal to the time rate of change of the incident optical momentum. When the surrounding medium is a transparent liquid, we showed the need to include the force of radiation on the body of the liquid in order to achieve consistency with momentum-based calculations. Division of the force between the solid and its surrounding liquid was examined in some detail, and three alternative methods of computing the "effective" force on the prism were suggested. Whether the effective force is in fact given by Eq. (22), Eq. (25), or Eq. (27) cannot be decided solely on the basis of theoretical considerations, but requires refined measurements of the force of radiation experienced by different solids immersed in a variety of compatible liquids.

## Acknowledgments

The authors are grateful to Ewan Wright and Pavel Polynkin for many helpful discussions. This work has been supported by the AFOSR contract F49620-02-1-0380 with the Joint Technology Office, by the *Office of Naval Research* MURI grant No. N00014-03-1-0793, and by the *National Science Foundation* STC Program under agreement DMR-0120967.